\documentclass[aip, apl,
 amsmath,amssymb,
 reprint,%
]{revtex4-1}

\usepackage{graphicx}
\usepackage{dcolumn}
\usepackage{bm}

\usepackage{hyperref}
\usepackage[utf8]{inputenc}
\usepackage[T1]{fontenc}
\usepackage{mathptmx}
\usepackage{etoolbox}
\usepackage{color}
\makeatletter
\def\@email#1#2{%
 \endgroup
 \patchcmd{\titleblock@produce}
  {\frontmatter@RRAPformat}
  {\frontmatter@RRAPformat{\produce@RRAP{*#1\href{mailto:#2}{#2}}}\frontmatter@RRAPformat}
  {}{}
}%
\makeatother
\begin{document}

\preprint{AIP/123-QED}

\title[Surface acoustic wave enabled all-optical determination of the interlayer elastic constants of van der Waals interface]{Surface acoustic wave enabled all-optical determination of the interlayer elastic constants of van der Waals interface}
\author{N.Yu. Frolov}
 
\author{A.Yu. Klokov}%
 \email{klokov@lebedev.ru}

\author{A. I. Sharkov}
 
\author{M.V. Pugachev}
\affiliation{
P.N. Lebedev Phyisical Institute of the Russian Academy of Sciences, 119991, Moscow, Russia}  
\author{A.Yu. Kuntsevich}
\affiliation{%
HSE University, Moscow, Russia 101000
}%
\altaffiliation[Also at ]{
P.N. Lebedev Phyisical Institute of the Russian Academy of Sciences, 119991, Moscow, Russia}

\date{\today}

\begin{abstract}
Understanding the properties of two-dimensional materials interfaces with the substrate is necessary for device applications. Surface acoustic wave propagation through the layered material flake on a substrate could provide unique information on the transverse rigidity of the flake-to-substrate interaction. We generate ultrasonic waves by a focused femtosecond laser pulse at the surface of the model system -- fused silica with h-BN flake transferred above. Using an all-optical spatially resolved pump-probe interferometric technique, we measure the spatial dependencies of the surface  vertical velocity profiles. Our measurements reveal the appearance of the surface acoustic wave dispersion in the hBN flake region compared to fused silica surface. Multilayer modeling allows us to gain access to longitudinal and shear elastic coupling constants $c^*_{33}$ and $c^*_{44}$ between hexagonal BN and substrate. 
\end{abstract}

\maketitle

The substrate is an inevitable, yet crucial, component for almost any possible applications of 2D materials and heterostructures. Van der Waals bond between the heterostructure and the substrate in both orthogonal and lateral directions is related to the mechanical stability of the structure and transferring the deformation between the substrate and the structure (straintronics\cite{miao2021straintronics}). 
Acoustic mismatch is responsible for finite heat conduction of the interfaces\cite{song2018two, zhang2020size}, a key limiting factor for low-temperature bolometrical\cite{miao2018graphene, kokkoniemi2020bolometer} and  thermoelectrical\cite{skoblin2018graphene}  applications of 2D materials. There are numerous surface acoustic wave-based device suggestions with 2D materials\cite{ZHOU2018389,LI2022113573,yoon2022mm}, where the role of the interface is crucial.

Despite a high demand for understanding the elastic parameters of the 2D material interfaces, they remain poorly investigated. 
Static measurements of the interface rigidity, e.g. with atomic force microscope tip, are difficult to interpret.

The elastic properties of 2D flakes could be explored from the dynamic effects, i.e., propagation of the surface acoustic wave (SAW). Interdigital transducer-based techniques are traditionally used for the acoustic probing. However combination of this technique with micrometer-size flakes appears to be tricky. Use of all-optical techniques, where both excitation and probing of sound is performed by optical pulses\cite{abi2024progress} is an explicit way to access the elastic parameters of 2D materials. 

Mathematically, rigidity defines the boundary condition for the strain and displacement at the interface.
In several previous picoacoustic studies with layered materials on substrates the longitudinal wave was laser-excited across the layers, that allowed to evaluate rigidity only in perpendicular direction \cite{greener2018coherent, greener2019high, klokov2022, wang2022revealing, wang2024optical}.
Surface acoustic wave (SAW) is different and more complicated because it involves both longitudinal and transversal components of the displacement and strain for layer and substrate in both vertical and lateral directions\cite{hess2002}. In this paper we use SAW as a tool to explore the properties of the van der Waals interface between prototypical layered material hexagonal boron nitride and SiO$_2$ substrate.

We excite SAW locally using repeated focused femtosecond pump pulses. SAW propagation is monitored by scanning the surface and precise measuring the change in the complex reflectivity coefficient with an infrared probe pulse.  It turns out that when passing through the hexagonal boron nitride flake located at the fused silica surface the wavefront acquires a strong dispersion. This dispersion is determined in particular by acoustically non-ideal flake-to-substrate contact. From the analysis of the dispersion one may determine previously inaccesible van der Waals bond parameters.

\begin{figure}
\includegraphics[width=0.5\textwidth]{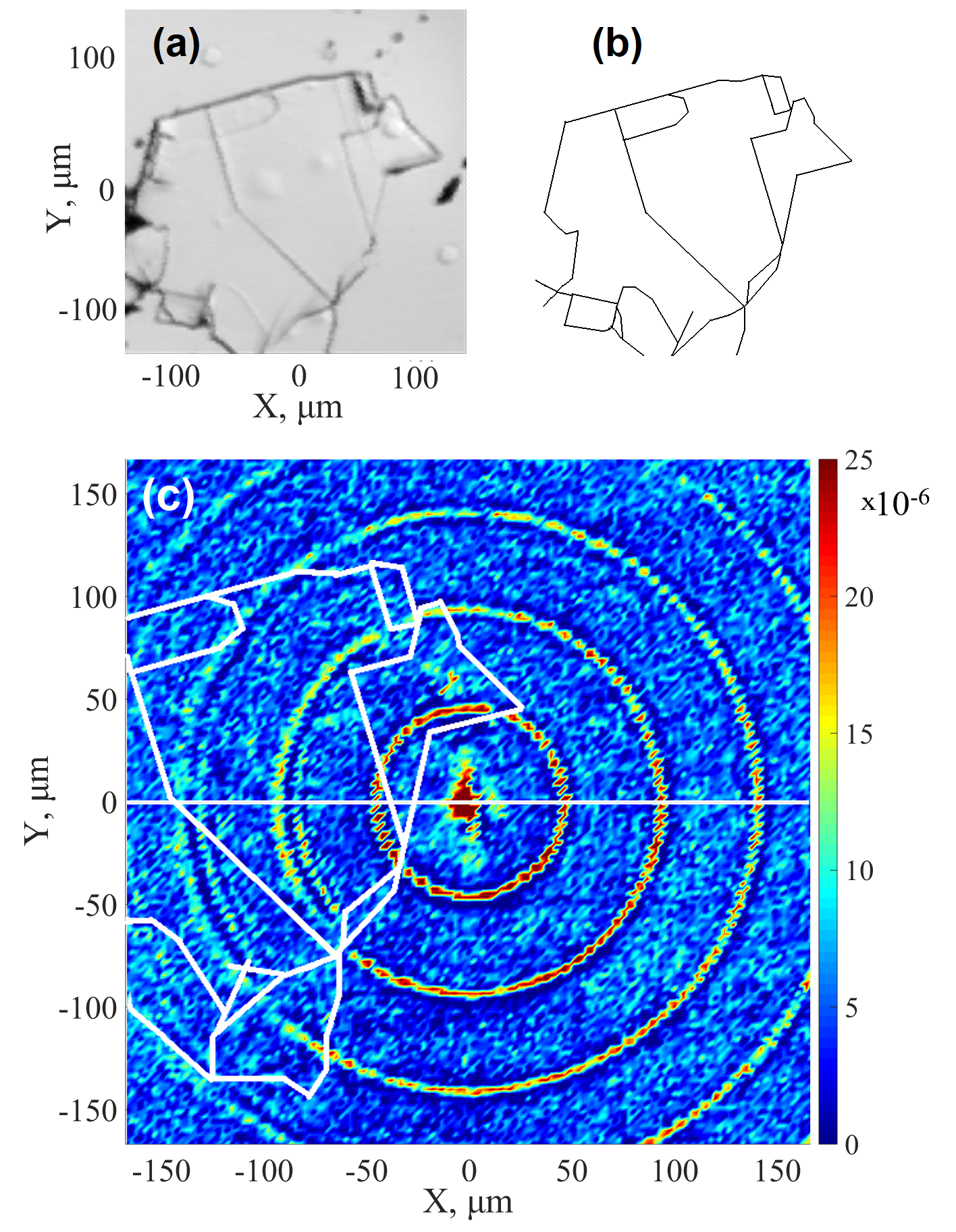}
\caption{(a) Reflectivity scan of the studeied sample. (b) hBN contour plot (c) Color map of the scanned simultaneous surface wavefield near the edge of the hBN flake.}
\label{ColorMap}
\end{figure}

Single crystal hBN from \href{https://2dsemiconductors.com/h-BN-large/}{2dsemiconductors.com} was first scotch-tape exfoliated onto Si/SiO$_2$ substrate, then characterized using atomic force microscopy, and transferred trough polypropylencarbonate \cite{martanov2020} onto a fused silica substrate with lithographically defined markers. To remove the organic residuals the sample was cleaned in acetone, isopropanol and deionized water and annealed in vacuum at 400 $^\circ$C for 3 hours. Then 30 nm transducer Al layer was e-beam evaporated atop. An image of the sample is shown in Fig.~\ref{ColorMap}a. The contour plot Fig.~\ref{ColorMap}b allows to identify large area regions with constant hBN thickness. The studied area had 600~nm thickness measured by atomic-force microscopy and interference microscopy.

The measurements were performed within two-color confocal scheme, see Refs.~[\onlinecite{tachizaki2006scanning},\onlinecite{klokov2022sensors}]. Pump pulses at second harmonic of Ti-Sapphire Coherent Mira Laser (repetition rate 76 MHz, pulse width 140 fs, pulse energy 2 nJ) were separated from the first harmonic probe pulses ($\lambda\approx 800$ nm) using dichroic filter. Both pump and probe were fed through the same micro-objective (NA=0.65, spot diameter $\sim$1.5~$\mu$m) to the sample. Probe pulse spot was scanned along the sample plane using motorized 4$f$-scanner. Sagnac interferometer was used to increase the sensitivity and detune from the low-frequency vibrations. The probe pulse splits into two with $\tau\approx 500$ ps delay. So the value $\Delta\phi=Im[(\delta R(t)-\delta R(t-\tau))/R]$ is measured, where $R()$ is the reflectivity at the moments of time $t$ and $t-\tau$, $\Delta\phi$ is proportional to the vertical velocity.

\begin{figure}
\includegraphics[width=0.5\textwidth]{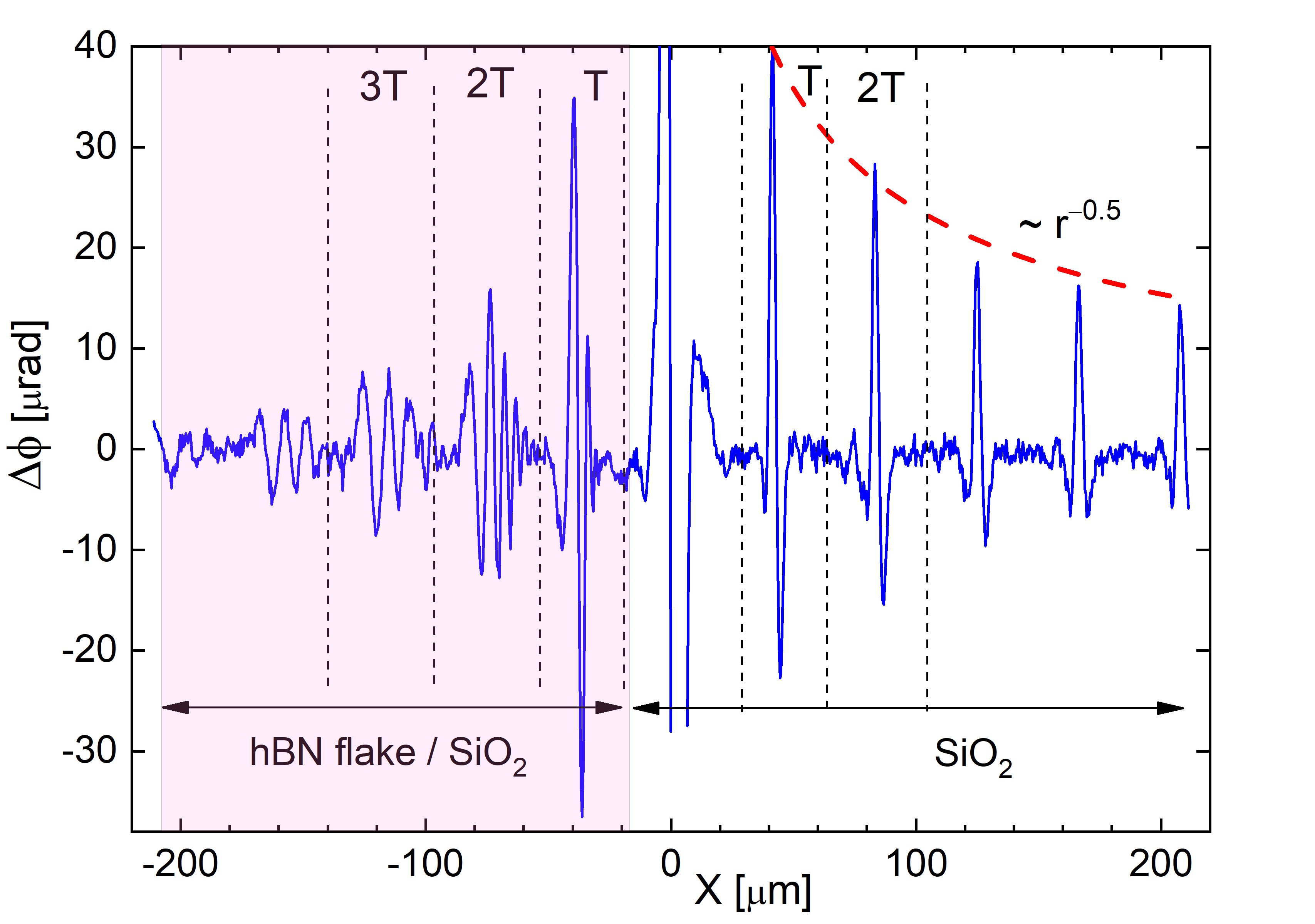}
\caption{Spatial dependence of the instant wavefield along the white horizontal line ($Y=0$) in Fig. \protect{\ref{ColorMap}}c.}
\label{PulseProfile}
\end{figure}

On a physical level laser pulse is absorbed mostly in top transducer layer forming a picosecond-scale strain pulse, essential part of which penetrates downwards to the substrate. Also some part of the initial picosecond pulse excites SAW. Probe pulse scans the displacement of the surface through complex reflectivity coefficient as a function of the probe position at a given delay time.  An example of such scan (reflectivity variation map) is shown in Fig.~\ref{ColorMap}c. Almost concentric rings correspond to repeating laser pulses (13 ns period). It is seen that as the pulse propagates along the hBN flake, its shape undergoes changes. A region along the horizontal $Y=0$ line was measured with high spatial resolution (0.5 $\mu$m) for the detailed analysis. The corresponding profile is shown in Fig.~\ref{PulseProfile}. The shape of the pulses propagating along SiO$_2$ remains visibly unchanged and  the amplitude of the pulses drops with radius $\propto 1/\sqrt{r}$ as it should be for the surface wave.
The pulses propagating along SiO$_2$/hBN change their shape considerably while propagating along hBN in contrast to SiO$_2$ propagation. This variation is due to dispersion of the surface acoustic wave. The major idea of this paper is the quantitative description of the shape change and evaluation of acoustic parameters via multi-layer model.

For quantitaive analysis of experimental data we take the 2D Fourier transform of the  SAW wavefield similarly to Ref. \cite{sugawara2002watching}. Axial symmetry of the problem makes the spatial spectrum dependent on $k=|\vec{k}|$, moreover this spectrum is a purely real function. Taking SAW decay into account one gets:
\begin{equation}
\Delta\phi(r,t)=\int \Phi(k,t)exp(\vec{k}\cdot\vec{r})d^2\vec{k},
\end{equation}
where spatial spectrum $\Phi(k,t)$ is 
\begin{equation}
\Phi(k,t)=|F_0(k)|e^{-\alpha(k)t}cos(\omega(k)t-\psi(k)).
\label{Eq2}
\end{equation}

Similarly to Ref. \cite{wright2002real} we consider SAW field as a sum of the independent pulses registered at moments of time $T-\delta T$, $2T-\delta T$, $3T-\delta T$, where $T$ is a repetition period of the laser, and $\delta T$ is certain unknown time of the propagation of SAW along the bare SiO$_2$ surface from the point of excitation to the hBN flake. As we show later, this time cancels out when dispersion of the SAW is determined from the data.

\begin{figure}
\includegraphics[width=0.5\textwidth]{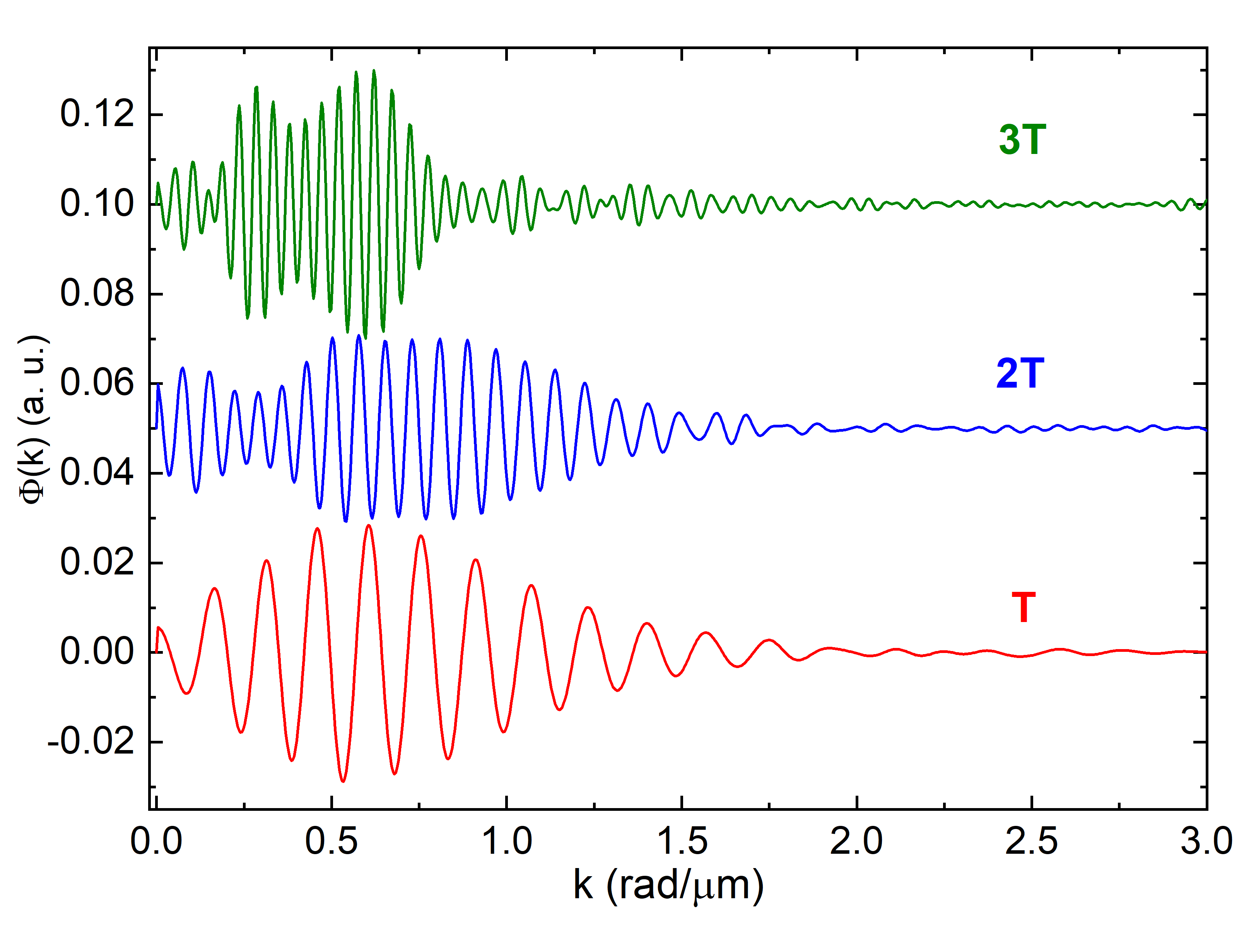}
\caption{
Measured spatial spectra $\Phi(k,t)$ of the SAW pulses on hBN/SiO$_2$ at $t=T$ (red), $t=2T$ (blue) and $t=3T$ (green). The spectra are shifted vertically for clarity.}
\label{SAWspectra}
\end{figure}

Fig.~\ref{SAWspectra} shows the spatial spectra of the separate SAW pulses from Fig.~\ref{PulseProfile} emitted by the laser pulses one, two and three repetition periods ahead the observation moment. In order to perform 2D Fourier transform 
we extrapolate the data of Fig.~\ref{PulseProfile} onto the whole $(X,Y)$ plane, i.e we assume that 
$\Delta\phi(x,y,t)$ is a function of $r=\sqrt{x^2+y^2}$ and $t$.

As seen from Fig.~\ref{SAWspectra} the spectra are fast oscillating wavepackets with a smooth envelope. The frequency of oscillations is $k$-dependent and is roughly proportional to the registration time ($T$, $2T$ or $3T$). The spectra at the moments $T$ and $2T$ span to $\sim$2 rad/$\mu$m. Using the SAW velocity for the SiO$_2$ substrate $v_{ph}$ (3.4 $\mu$/ns ) we get the time spectra upper range estimate as $\sim$1 GHz. Spatial spectrum at $3T$ is 1.5 times narrower. We relate this to the high frequency components scattering or damping by structural defects.

The dispersion is nothing but $\omega(k)$ dependence that has to be determined from the experimental data. A method, based on division of the spectra taken at different moments of time  to determine $\omega(k)$ dependence was suggested e.g. in Ref.~[\onlinecite{wright2002real}]. Unfortunately this method is unstable and drastically sensitive to noise and errors.
We use an alternative approach to determine  $\omega(k)t=kv_{ph}(k)t$, where $v_{ph}$ is $k$-dependent phase velocity. Within the theory of signals the product $v_{ph}t$ is analogous to frequency, whereas $k$ is analogous to time. Our goal is to determine $v_{ph}t$ product. 

A standard method for such problems is to use the  Hilbert transform, i.e. formation of “analytic signal”\cite{born1980basic} $V(k)e^{i\varphi(k)}$. The spectrum of analytic signal is two times higher than that for initial signal for positive frequencies and equals to zero at negative frequencies. Usual practice of its calculation is multiplication of the Fourier transform of the initial signal by 2 for positive frequencies and by 0 - for negative frequencies. Than the inverse Fourier transform is performed. The module of the analytical signal produces an envelope whereas its argument is nothing but the instant phase. 
Taking Eq.~\ref{Eq2} into account the phase $\varphi(k,t)$ is expressed as follows:
\begin{equation}
\varphi(k,t)=\omega(k)t-\psi(k).
\end{equation}

Thus we suggest the following algorithm to determine the phase velocity:
\begin{enumerate}
\item{We perform a 2D Fourier transform of the SAW pulses to determine a space spectrum $\Phi(k,T-\delta T)$ of the first pulse and $\Phi(k,2T-\delta T)$ for the second pulse.}
\item{“Analytic” signals are formed from these space spectra}
\item{Arguments of the analytic signals from the second and the first pulses} are subtracted from each other that gives explicitly $k\cdot v_{ph}(k)(2T-\delta T)-kv_{ph}(k)(T-\delta T)=kv_{ph}(k)T$
\item{Dividing the argument difference by $kT$ we obtain phase velocity $v_{ph}(k)$.}
\end{enumerate}

\begin{figure}
\includegraphics[width=0.5\textwidth]{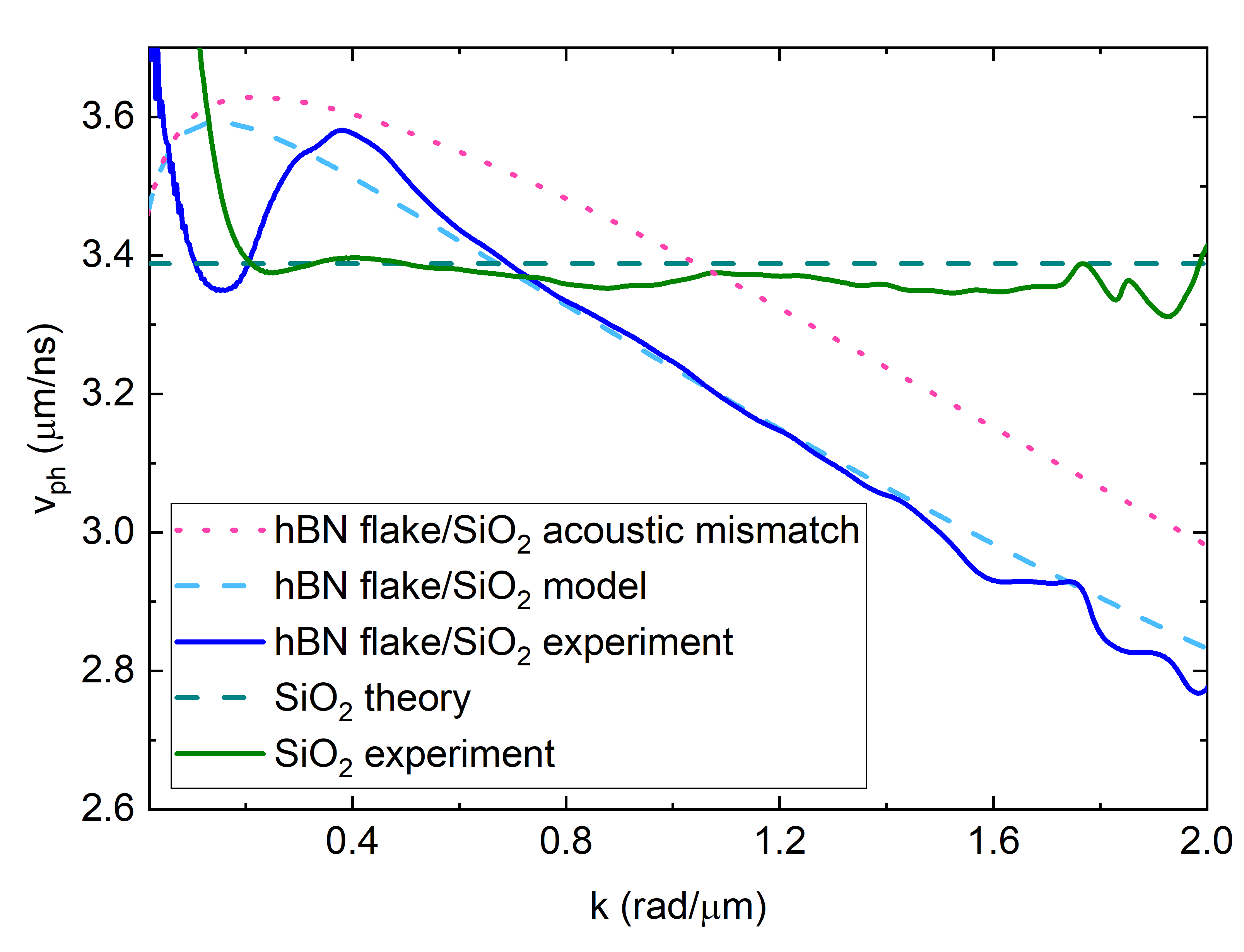}
\caption{Wave vector dependence of the phase velocity. Green colors - bare SiO$_2$ surface; Blue colors - SiO$_2$/hBN layered structure. Solid lines denote experimental data, dashed lines are theory (constant velocity for SiO$_2$ and result of multilayered modelling for SiO$_2$/hBN). Red dotted curve is a simulation of the SiO$_2$/hBN SAW dispersion within the acoustic mismatch model.}
\label{Dispersion}
\end{figure}

The phase velocity is shown in Fig. \ref{Dispersion} by solid blue line. There is a clear tendency to decrease of the phase velocity with $k$.  In the obtained dispersion curves, errors in the small wave vector region are introduced by splitting the total signal into individual pulses. Similar low-$k$ features emerged also in Ref.~[\onlinecite{sugawara2002watching}](in Fig.~1d therein the amplitude does not tend to zero as $k=0$). 

The errors in the large wave vector values region are caused by a decrease in the signal-to-noise ratio.
High frequency cut-off $k<2$~rad/$\mu$m corresponds to frequencies $\sim 1$~GHz and is limited by the excitation spot spatial spectrum (i.e. spot size) and probe spatial resolution.

To demonstrate the applicability of our method we present in Fig.~\ref{Dispersion} for comparison the dispersion for bare SiO$_2$ surface (solid green line). Except for low-$k$ feature, there is no dispersion, as expected for Rayleigh wave at the flat solid surface (dashed green line). The dispersion is therefore due to the flake and signifies that the flake and SiO$_2$ are acoustically well-coupled.

Multilayered elastic models should quantify the SAW dispersion in the composite structures. 
Experiments on optical SAW detection in thin films and layered structures\cite{lehmann2002y, philip2003, haim2011a, salenbien2011laser, sermeus2014, heczko2018,liu2021, makowski2024} often consider an ideal interface. It is called acoustic mismatch boundary condition and means continuous stress and displacement across the interface. This approach seems natural when the unknown parameters of the studied films are to be evaluated. If the parameters of the film and the substrate are known and acoustic mismatch does not describe the SAW dispersion, non-ideal interfaces should be assumed. 

Crystalline symmetry allows 5 nonzero components in the elastic tensor of hBN, and all of them are known\cite{bosak2006elasticity} ($c_{11}=c_{22}=811$~GPa, $c_{33}=$~27 GPa, $c_{44}=c_{55}=7.7$~GPa, $c_{12}=169$~GPa, $c_{13}=0$, $c_{66}=(c_{11}-c_{12}$)/2); isotropic semi-infinite substrate has two relevant components  $c_{33}=77.5$~GPa and $c_{44}=30.8$~GPa. The densities of both hBN and SiO$_2$ are equal to 2200~kg/m$^3$. The substrate parameters explain perfectly the observed dispersionless SAW on bare SiO$_2$ ( compare dahed and solid green lines  in Fig.~\ref{Dispersion}). The result of the acoustic mismatch modeling for hBN/SiO$_2$ structure is shown by dotted red line in Fig.~\ref{Dispersion}. It is clear that the agreement with experiment is poor and it could not be improved by reasonable variation of the material parameters. This observation suggests the necessity to introduce a non-ideal acoustic coupling at the interface. 

Previous longitudinal wave picoacoustic experiments with 2D materials\cite{greener2018coherent,klokov2022,wang2022revealing} demonstrated distributed spring layer at the van der Waals interfaces. The interlayer interaction in the multilayered models is often described as a layer with small thickness and mass and some elastic constants\cite{du2010evaluation}. 
  By taking zero density limit one obtains the following boundary conditions between the layer and the substrate:
\begin{equation}
   U^L_z=U^S_z+\sigma^L_{zz}/c^*_{33} 
\end{equation}
\begin{equation}
   \sigma^L_{zz}=\sigma^S_{zz} 
\end{equation}
\begin{equation}
   U^L_x=U^S_x+\sigma^L_{xz}/c^*_{44} 
\end{equation}
\begin{equation}
   \sigma^L_{xz}=\sigma^S_{xz} 
\end{equation}
Here $\sigma$ is the the stress tensor,  $U$ is the displacement vector, $S$ denotes substrate and $L$ denotes layer, and $c^*_{33}$ and $c^*_{44}$ are the longitudinal and lateral elastic parameters quantifying the interlayer interaction.

Elastic constant $c_{33}^*$ and $c_{44}^*$ serve as adjustable parameters. The results of the calculation within this model with $c^*_{33}=10^{19}$~N/m$^3$ and $c^*_{44}=2.7\cdot 10^{16}$~N/m$^3$ are shown by dashed blue line in Fig.~\ref{Dispersion} . We also take into account Al transducer as a $\delta$-function distributed mass for all calculations, but it did not change the results visibly.  We are not aware of any alternative measurements of $c^*_{44}$ constant of the VdW interface VdW heterostructures.

The value $c^*_{33}=1.0\cdot10^{19}$~N/m$^3$ agrees reasonably with the the results of the longitudinal acoustic waves experiments.
Ref.~[\onlinecite{greener2018coherent}] reports $c_{33}$ well below $10^{20}$~N/m$^3$ for InSe/Al$_2$O$_3$ interface. In Ref.~[\onlinecite{klokov2022}] $c_{33}\approx 8\cdot 10^{18}$~N/m$^3$ for hBN on Al$_2$O$_3$ and similar values for the other interfaces are determined. Ref.~[\onlinecite{wang2022revealing}] reports $c_{33}\approx 1.5\cdot 10^{19}$~N/m$^3$ for MoS$_2$ on Al$_2$O$_3$ within higher frequency experiments. This agreement points to the correctness of the used multi-layer SAW model.


The method could be extended to determination of the elastic constants of heterointefaces in the multilayered structures. This will require more sophisticated data processing since the problem becomes ill-conditioned\cite{salenbien2011laser}.  
 An interesting optical alternative for the SAW dispersion measurement is the transient grating method\cite{rogers2000optical}, that would be promissing to implement in 2D materials. Concerning the applications, Ref.~[\onlinecite{yoon2022mm}] suggests using sound waves in hBN for acoustoelectronic filters. Taking hBN-to-susbtrate acoustic coupling into account should crucially determine the characteristics of such devices.

In conclusion, we applied an all-optical scanning pump-probe technique to study the layered van der Waals material flakes on the substrate by means of surface acoustic waves. It is shown that when a layer of hexagonal boron nitride is placed onto fused silica surface, the surface acoustic wave pulses become distorted due to dispersion of the SAW and pulse shape variation allows to evaluate the dispersion. As follows from the modeling, the dispersion is determined by longitudinal and transversal interlayer coupling constants, and the latter could be found. The method thus provides a unique and demanded information about the mechanical properties of the van der Waals interfaces.

\begin{acknowledgments}
The work is supported by Russian Science Foundation (Grant No, 24-72-00132). Sample fabrication was performed at the Shared Facility Center of the P.N. Lebedev Physical Institute. A.Yu. Kuntsevich thanks Basic Research program of the HSE for support. 
\end{acknowledgments}

\section*{Data Availability Statement}

The data that support the findings of this study are available from the corresponding author upon reasonable request.

\nocite{*}
\bibliography{aipsamp}

@PREAMBLE{
 "\providecommand{\noopsort}[1]{}" 
 # "\providecommand{\singleletter}[1]{#1}%" 
}

@article{klokov2022sensors,
  title={Optical excitation of converging surface acoustic waves in the gigahertz range on silicon},
  author={Klokov, Andrey Y and Krivobok, Vladimir S and Sharkov, Andrey I and Frolov, Nikolay Y},
  journal={Sensors},
  volume={22},
  number={3},
  pages={870},
  year={2022},
  publisher={MDPI}
}

@article{sugawara2002watching,
  title={Watching ripples on crystals},
  author={Sugawara, Yl and Wright, OB and Matsuda, O and Takigahira, M and Tanaka, Yl and Tamura, S and Gusev, VE},
  journal={Physical review letters},
  volume={88},
  number={18},
  pages={185504},
  year={2002},
  publisher={APS}
}

@article{du2010evaluation,
  title={Evaluation of film adhesion to substrates by means of surface acoustic wave dispersion},
  author={Du, Jikai and Tittmann, Bernhard R and Ju, Hyeong Sick},
  journal={Thin Solid Films},
  volume={518},
  number={20},
  pages={5786--5795},
  year={2010},
  publisher={Elsevier}
}

@article{born1980basic,
  title={Basic properties of the electromagnetic field},
  author={Born, Max and Wolf, Emil},
  journal={Principles of optics},
  volume={44},
  pages={1--70},
  year={1980},
  publisher={Elsevier}
}

@article{greener2018coherent,
  title={Coherent acoustic phonons in van der Waals nanolayers and heterostructures},
  author={Greener, Jake DG and Akimov, Andrey V and Gusev, VE and Kudrynskyi, ZR and Beton, Peter H and Kovalyuk, Zakhar D and Taniguchi, Takashi and Watanabe, Kenji and Kent, AJ and Patan{\`e}, Amalia},
  journal={Physical Review B},
  volume={98},
  number={7},
  pages={075408},
  year={2018},
  publisher={APS}
}

@article{rogers2000optical,
  title={Optical generation and characterization of acoustic waves in thin films: Fundamentals and applications},
  author={Rogers, John A and Maznev, Alex A and Banet, Matthew J and Nelson, Keith A},
  journal={Annual Review of Materials Science},
  volume={30},
  number={1},
  pages={117--157},
  year={2000},
  publisher={Annual Reviews 4139 El Camino Way, PO Box 10139, Palo Alto, CA 94303-0139, USA}
}

@article{bosak2006elasticity,
  title={Elasticity of hexagonal boron nitride: Inelastic x-ray scattering measurements},
  author={Bosak, Alexey and Serrano, Jorge and Krisch, Michael and Watanabe, Kenji and Taniguchi, Takashi and Kanda, Hisao},
  journal={Physical Review B—Condensed Matter and Materials Physics},
  volume={73},
  number={4},
  pages={041402},
  year={2006},
  publisher={APS}
}

@article{wright2002real,
  title={Real-time imaging and dispersion of surface phonons in isotropic and anisotropic materials},
  author={Wright, OB and Sugawara, Y and Matsuda, O and Takigahira, M and Tanaka, Y and Tamura, S and Gusev, VE},
  journal={Physica B: Condensed Matter},
  volume={316},
  pages={29--34},
  year={2002},
  publisher={Elsevier}
}

@article{wang2022revealing,
  title={Revealing the interlayer van der Waals coupling of bi-layer and tri-layer MoS$_2$ using terahertz coherent phonon spectroscopy},
  author={Wang, Peng-Jui and Tsai, Po-Cheng and Yang, Zih-Sian and Lin, Shih-Yen and Sun, Chi-Kuang},
  journal={Photoacoustics},
  volume={28},
  pages={100412},
  year={2022},
  publisher={Elsevier}
}

@article{wang2024optical,
  title={Optical quantification of the weak van der Waals coupling between multilayer antimonene and bilayer MoS$_2$ using ultrafast coherent vibration spectroscopy},
  author={Wang, Peng-Jui and Yang, Zih-Sian and Chang, Che-Jia and Lin, Shih-Yen and Sun, Chi-Kuang},
  journal={The Journal of Chemical Physics},
  volume={161},
  number={9},
  pages={094704},
year={2024},
  publisher={AIP Publishing}
}

@article{skoblin2018graphene,
  title={Graphene bolometer with thermoelectric readout and capacitive coupling to an antenna},
  author={Skoblin, Grigory and Sun, Jie and Yurgens, August},
  journal={Applied Physics Letters},
  volume={112},
  number={6},
  year={2018},
  publisher={AIP Publishing}
}

@article{miao2021straintronics,
  title={Straintronics with van der Waals materials},
  author={Miao, Feng and Liang, Shi-Jun and Cheng, Bin},
  journal={NPJ Quantum Materials},
  volume={6},
  number={1},
  pages={59},
  year={2021},
  publisher={Nature Publishing Group UK London}
}

@article{martanov2020,
  title={Making van der Waals heterostructures assembly accessible to everyone},
  author={Martanov, Sergey G and Zhurbina, Natalia K and Pugachev, Mikhail V and Duleba, Aliaksandr I and Akmaev, Mark A and Belykh, Vasilii V and Kuntsevich, Aleksandr Y},
  journal={Nanomaterials},
  volume={10},
  number={11},
  pages={2305},
  year={2020},
  publisher={MDPI}
}

@article{abi2024progress,
  title={Progress in laser ultrasonics evaluation of micro-and nanoscale interfacial mechanics},
  author={Abi Ghanem, Maroun and Dehoux, Thomas},
  journal={Applied Physics Reviews},
  volume={11},
  number={4},
  year={2024},
  publisher={AIP Publishing}
}

@article{hess2002,
  title={Surface acoustic waves in materials science},
  author={Hess, Peter},
  journal={Physics Today},
  volume={55},
  number={3},
  pages={42--47},
  year={2002},
  publisher={AIP Publishing}
}

@article{lehmann2002y,
  title={Young’s modulus and density of nanocrystalline cubic boron nitride films determined by dispersion of surface acoustic waves},
  author={Lehmann, G and Hess, P and Weissmantel, S and Reisse, G and Scheible, P and Lunk, A},
  journal={Applied Physics A},
  volume={74},
  number={1},
  pages={41--45},
  year={2002},
  publisher={Springer}
}

@article{ZHOU2018389,
title = {2-Dimentional photoconductive MoS2 nanosheets using in surface acoustic wave resonators for ultraviolet light sensing},
journal = {Sensors and Actuators A: Physical},
volume = {271},
pages = {389-397},
year = {2018},
issn = {0924-4247},
doi = {https://doi.org/10.1016/j.sna.2017.12.007},
url = {https://www.sciencedirect.com/science/article/pii/S092442471731484X},
author = {Peng Zhou and Changsong Chen and Xiang Wang and Baofa Hu and Haisheng San}
}

@article{LI2022113573,
title = {A high-sensitivity MoS2/graphene oxide nanocomposite humidity sensor based on surface acoustic wave},
journal = {Sensors and Actuators A: Physical},
volume = {341},
pages = {113573},
year = {2022},
issn = {0924-4247},
doi = {https://doi.org/10.1016/j.sna.2022.113573},
url = {https://www.sciencedirect.com/science/article/pii/S0924424722002114},
author = {Xiangrong Li and Qiulin Tan and Li Qin and Lei Zhang and Xiaorui Liang and Xiawen Yan}
}

@article{philip2003,
  title={Elastic, mechanical, and thermal properties of nanocrystalline diamond films},
  author={Philip, J and Hess, P and Feygelson, T and Butler, JE and Chattopadhyay, S and Chen, KH and Chen, LC},
  journal={Journal of Applied Physics},
  volume={93},
  number={4},
  pages={2164--2171},
  year={2003},
  publisher={American Institute of Physics}
}

@article{liu2021,
  title={Estimation of silicon wafer coating thickness using ultrasound generated by femtosecond laser},
  author={Liu, Peipei and Yi, Kiyoon and Sohn, Hoon},
  journal={Journal of Nondestructive Evaluation, Diagnostics and Prognostics of Engineering Systems},
  volume={4},
  number={1},
  pages={011005},
  year={2021},
  publisher={American Society of Mechanical Engineers}
}

@article{makowski2024,
  title={Surface acoustic wave spectroscopy for non-destructive coating and bulk characterization at temperatures up to 600° C enabled by piezoelectric aluminum nitride coated sensor},
  author={Makowski, Stefan and Zawischa, Martin and Schneider, Dieter and Barth, Stephan and Schettler, Sebastian and Hoang, Thanh-Tung and Bartzsch, Hagen and Zimmermann, Martina},
  journal={Surface and Interface Analysis},
  volume={56},
  number={5},
  pages={319--332},
  year={2024},
  publisher={Wiley Online Library}
}

@article{heczko2018,
  title={Temperature dependence of elastic properties in austenite and martensite of Ni-Mn-Ga epitaxial films},
  author={Heczko, Oleg and Seiner, Hanu{\v{s}} and Stoklasov{\'a}, Pavla and Sedl{\'a}k, Petr and Sermeus, Jan and Glorieux, Christ and Backen, Anja and F{\"a}hler, Sebastian and Landa, Michal},
  journal={Acta Materialia},
  volume={145},
  pages={298--305},
  year={2018},
  publisher={Elsevier}
}

@article{tachizaki2006scanning,
  title={Scanning ultrafast Sagnac interferometry for imaging two-dimensional surface wave propagation},
  author={Tachizaki, Takehiro and Muroya, Toshihiro and Matsuda, Osamu and Sugawara, Yoshihiro and Hurley, David H and Wright, Oliver B},
  journal={Review of Scientific Instruments},
  volume={77},
  number={4},
 page={043713}, 
year={2006},
  publisher={AIP Publishing}
}

@article{haim2011a,
  title={Elastic characterization of Au thin films utilizing laser induced acoustic Rayleigh waves},
  author={Haim, A and Bar-Ad, S and Azoulay, A},
  journal={Journal of Physics: Conference Series},
  volume={278},
  number={1},
  pages={012005},
  year={2011},
  organization={IOP Publishing}
}

@article{sermeus2014,
  title={Determination of elastic properties of a MnO2 coating by surface acoustic wave velocity dispersion analysis},
  author={Sermeus, Jan and Sinha, R and Vanstreels, Kris and Vereecken, PM and Glorieux, Christ},
  journal={Journal of Applied Physics},
  volume={116},
  number={2},
  year={2014},
  page={023503},
  publisher={AIP Publishing}
}

@article{salenbien2011laser,
  title={Laser-based surface acoustic wave dispersion spectroscopy for extraction of thicknesses, depth, and elastic parameters of a subsurface layer: Feasibility study on intermetallic layer structure in integrated circuit solder joint},
  author={Salenbien, Robbe and C{\^o}te, R and Goossens, J and Limaye, Paresh and Labie, Riet and Glorieux, Christ},
  journal={Journal of Applied Physics},
  volume={109},
  number={9},
  year={2011},
  pages={093104},
  publisher={AIP Publishing}
}

@article{kokkoniemi2020bolometer,
  title={Bolometer operating at the threshold for circuit quantum electrodynamics},
  author={Kokkoniemi, Roope and Girard, J-P and Hazra, Dibyendu and Laitinen, Antti and Govenius, Joonas and Lake, RE and Sallinen, Iiro and Vesterinen, Visa and Partanen, Matti and Tan, JY and others},
  journal={Nature},
  volume={586},
  number={7827},
  pages={47--51},
  year={2020},
  publisher={Nature Publishing Group}
}

@article{miao2018graphene,
  title={A graphene-based terahertz hot electron bolometer with Johnson noise readout},
  author={Miao, W and Gao, H and Wang, Z and Zhang, W and Ren, Y and Zhou, KM and Shi, SC and Yu, C and He, ZZ and Liu, QB and others},
  journal={Journal of Low Temperature Physics},
  volume={193},
  number={3},
  pages={387--392},
  year={2018},
  publisher={Springer}
}

@article{song2018two,
  title={Two-dimensional materials for thermal management applications},
  author={Song, Houfu and Liu, Jiaman and Liu, Bilu and Wu, Junqiao and Cheng, Hui-Ming and Kang, Feiyu},
  journal={Joule},
  volume={2},
  number={3},
  pages={442--463},
  year={2018},
  publisher={Elsevier}
}

@article{zhang2020size,
  title={Size-dependent phononic thermal transport in low-dimensional nanomaterials},
  author={Zhang, Zhongwei and Ouyang, Yulou and Cheng, Yuan and Chen, Jie and Li, Nianbei and Zhang, Gang},
  journal={Physics Reports},
  volume={860},
  pages={1--26},
  year={2020},
  publisher={Elsevier}
}

@article{yoon2022mm,
  title={mm-band surface acoustic wave devices utilizing two-dimensional boron nitride},
  author={Yoon, Seok Hyun and Baek, Chang-Ki and Kong, Byoung Don},
  journal={Scientific Reports},
  volume={12},
  number={1},
  pages={20578},
  year={2022},
  publisher={Nature Publishing Group UK London}
}

@article{greener2019high,
  title={High-frequency elastic coupling at the interface of van der Waals nanolayers imaged by picosecond ultrasonics},
  author={Greener, Jake DG and de Lima Savi, Elton and Akimov, Andrey V and Raetz, Samuel and Kudrynskyi, Zakhar and Kovalyuk, Zakhar D and Chigarev, Nikolay and Kent, Anthony and Patane, Amalia and Gusev, V.},
  journal={ACS nano},
  volume={13},
  number={10},
  pages={11530--11537},
  year={2019},
  publisher={ACS Publications}
}

@article{klokov2022,
  title={3D Hypersound microscopy of van der Waals heterostructures},
  author={Klokov, Andrey Yu and Frolov, Nikolay Yu and Sharkov, Andrey I and Nikolaev, Sergey N and Chernopitssky, Maxim A and Chentsov, Semen I and Pugachev, Mikhail V and Duleba, Aliaksandr I and Shupletsov, Alexey V and Krivobok, Vladimir S and others},
  journal={Nano Letters},
  volume={22},
  number={5},
  pages={2070--2076},
  year={2022},
  publisher={ACS Publications}
}

\end{document}